\documentclass[prc,preprint,showpacs,showkeys,
  tightenlines,
  amsfonts,nofootinbib]{revtex4}

\usepackage{graphicx}  %
\usepackage{bm}  %
\usepackage{amssymb}

\begin{document}

%

% general definitions

\newcommand{\beq}{\begin{equation}}
\newcommand{\eeq}{\end{equation}}
\newcommand{\bea}{\begin{eqnarray}}
\newcommand{\eea}{\end{eqnarray}}
\newcommand{\ben}{\begin{eqnarray*}}
\newcommand{\een}{\end{eqnarray*}}

\newcommand{\simlt}{\stackrel{<}{{}_\sim}}
\newcommand{\simgt}{\stackrel{>}{{}_\sim}}
\newcommand{\sing}{$^1\!S_0$ }
\newcommand{\btau}{\mbox{\boldmath$\tau$}}
\newcommand{\bsig}{\mbox{\boldmath$\sigma$}}

\newcommand{\dt}{\partial_t}

\newcommand{\kf}{k_{\rm F}}
\newcommand{\wt}{\widetilde}
\newcommand{\kt}{\widetilde k}
\newcommand{\pt}{\widetilde p}
\newcommand{\qt}{\widetilde q}
\newcommand{\wh}{\widehat}
\newcommand{\dens}{\rho}
\newcommand{\edens}{{\cal E}}
\newcommand{\order}[1]{{\cal O}(#1)}

\newcommand{\psihat}{\widehat\psi}
\newcommand{\xvec}{{\bf x}}
\newcommand{\dagphan}{{\phantom{\dagger}}}
\newcommand{\kvec}{{\bf k}}
\newcommand{\kpvec}{{\bf k}'}
\newcommand{\ak}{a^\dagphan_\kvec}
\newcommand{\akdag}{a^\dagger_\kvec}
\newcommand{\akv}[1]{a^\dagphan_{\kvec_{#1}}}
\newcommand{\akdagv}[1]{a^\dagger_{\kvec_{#1}}}
\newcommand{\akp}{a^\dagphan_{\kvec'}}
\newcommand{\akpdag}{a^\dagger_{\kvec'}}
\newcommand{\akpv}[1]{a^\dagphan_{\kvec'_{#1}}}
\newcommand{\akpdagv}[1]{a^\dagger_{\kvec'_{#1}}}

\def\vec#1{{\bf #1}}

\newcommand{\nab}{\overrightarrow{\nabla}}
\newcommand{\nabsq}{\overrightarrow{\nabla}^{2}\!}
\newcommand{\nabsqx}{\overrightarrow{\nabla}_x^{2}\!}
\newcommand{\nabl}{\overleftarrow{\nabla}}
\newcommand{\galnab}{\tensor{\nabla}}
\newcommand{\psid}{{\psi^\dagger}}
\newcommand{\psidal}{{\psi^\dagger_\alpha}}
\newcommand{\psidbe}{{\psi^\dagger_\beta}}
\newcommand{\idt}{{i\partial_t}}
\newcommand{\Sthree}{{\delta_{11'}(\delta_{22'}\delta_{33'}%
        -\delta_{23'}\delta_{32'})%
        +\delta_{12'}(\delta_{23'}\delta_{31'}-\delta_{21'}\delta_{33'})%
        +\delta_{13'}(\delta_{21'}\delta_{32'}-\delta_{22'}\delta_{31'})}}
\newcommand{\Stwo}{{\delta_{11'}\delta_{22'}-\delta_{12'}\delta_{21'}}}
\newcommand{\Left}{{\cal L}}
\newcommand{\Tr}{{\rm Tr}}

\newcommand{\h}{\hfil}
\newcommand{\be}{\begin{enumerate}}
\newcommand{\ee}{\end{enumerate}}
\newcommand{\I}{\item}

%
%
%
% End: Simple substitution macros
%

% uncomment \draft to have PACS numbers appear

\title{Effective Field Theory for Fermi Systems
       in a large $N$ expansion}

\author{R.J. Furnstahl}\email{furnstahl.1@osu.edu}
\author{H.-W. Hammer}\thanks{Address after October 1, 2002: Institut 
  f\"ur Theoretische Physik,
  Karl-Franzens-Universt\"at Graz, A-8010 Graz, Austria} 
%\email{hammer@mps.ohio-state.edu}
\affiliation{Department of Physics,
         The Ohio State University, Columbus, OH\ 43210}

%
%\date{\today}
\date{August, 2002}

\begin{abstract}
A system of fermions with short-range interactions at finite density 
is studied using the framework of effective field theory.
The effective action formalism for fermions with auxiliary fields 
leads to a loop expansion in which particle-hole bubbles are resummed to
all orders.  For spin-independent interactions, the loop expansion is 
equivalent to a systematic expansion in $1/N$,
where ``$N$'' is the spin-isospin degeneracy $g$. 
Numerical results at next-to-leading order are presented
and the connection to the Bose limit
of this system is elucidated.
\end{abstract}

\smallskip
\pacs{11.10.-z, 11.15.Pg, 05.30.Fk, 05.30.Jp}
\keywords{Effective field theory, effective action, large-$N$ expansion,
universality}
\maketitle

\section{Introduction}
\label{sec:introduction}

Effective field theory (EFT) provides a powerful framework to study
low-energy phenomena in a model-independent way by
exploiting the separation of scales in physical 
systems \cite{Weinberg79,LEPAGE89,KAPLAN95}.
Only low-energy (or long-range) degrees of freedom are included
explicitly, with the short-range physics 
parametrized in terms of the most general
local (contact) interactions.  Using renormalization,
the influence of high-energy states on low-energy observables
is captured in a small number of constants.
Thus, the EFT describes universal low-energy physics independent of
detailed assumptions about the high-energy dynamics.
Recent applications of EFT methods to nuclear physics
have made steady progress in the
two- and three-nucleon sectors \cite{EFT98,EFT99,Birareview,BEANE99,%
Bedaque:2002mn}.
In this paper, we continue a complementary program to apply EFT
methods to  many-fermion systems, with the ultimate goal of describing
nuclei and nuclear matter.

We showed the promise of using effective field theory 
for the nuclear many-body problem in Ref.~\cite{HAMMER00}, where  
a dilute, uniform gas of fermions interacting via short-range
interactions (which could be highly singular potentials such as hard
cores) was analyzed.
In general, this problem is nonperturbative in the potential, so 
the conventional diagrammatic treatment sums particle-particle 
ladder diagrams to all orders and thereby replaces the bare interaction by a
$K$ matrix \cite{FETTER71,BISHOP73}.  
Each $K$ matrix is in turn replaced by an effective range expansion
in momentum, and in the end  
one finds a perturbative expansion of the energy per particle
[see Eq.~(\ref{eq:edkf7})] in terms of 
the Fermi momentum $\kf$ times the $S$-wave scattering length $a_s$ 
(and other
effective range parameters in higher orders).

In contrast,
the EFT approach is  more direct, transparent, and systematic.  
The EFT automatically recasts the problem in the form of a perturbative
Fermi momentum expansion.
The freedom to use different regulators and renormalization schemes
was exploited by choosing dimensional regularization with minimal
subtraction (DR/MS), for which the dependence on dimensional
scales factors cleanly into $\kf$ dependence solely from the loop integrals
and dependence on effective range parameters solely in the
coefficients. 
This allowed power counting by simple
dimensional analysis and manifested the
universal nature of the low-order contributions.

Power counting refers to a procedure that identifies the contributions
to a given order in an EFT expansion; this is an essential feature of the
EFT approach. 
In the dilute case, the power counting with DR/MS 
is particularly simple: each
diagram contributes at a single order in $\kf$
and there are a finite number of
diagrams at each order.
This renormalization scheme also simplified a
renormalization group analysis to identify logarithmic divergences
that lead to nonanalytic (logarithmic) terms in the $\kf$ 
expansion of the energy \cite{Braaten97,HAMMER00}.

Thus, the application of EFT methods to the dilute Fermi system exhibits
a consistent organization of many-body corrections, with reliable error
estimates, and insight into the analytic structure of observables.
EFT provides a model-independent description of finite-density
observables in terms of parameters that can be fixed from
scattering in the vacuum or from a subset of finite-density properties.
The \emph{universal} nature of the expansion is also a key feature; 
\emph{any} underlying potential, probed at long distance, is
reproduced by the same form and the differences lie only in the low-energy
constants.
This perturbative analysis
is directly applicable to systems of trapped atoms 
and provides a controlled theoretical laboratory for studying issues 
such as the status of
occupation numbers as observables \cite{Furnstahl:2001xq}
or the nature of the Coester line \cite{FURNSTAHL01}.

However, bound nuclear systems and the most interesting phenomena
in atomic systems require a \emph{nonperturbative} EFT treatment.
The most  conspicuous features of nuclear interactions leading to
nonperturbative physics are the large $S$-wave scattering lengths. 
The analysis in Ref.~\cite{HAMMER00} assumes that the 
(spin-averaged) scattering
length is of natural size (e.g., of order the range of the underlying
interaction).
When the scattering length is large, the expansion applies only at
very small $\kf$, and  an alternative power counting is needed to 
extend the EFT to larger densities.
This power counting, developed in Refs.~\cite{vanKolck:1998bw} and
\cite{ksw} for two-particle scattering in free space, prescribes that 
leading order for large $a_s$ must sum
\emph{all} diagrams with non-derivative four-fermion ($C_0$) vertices
(although in higher orders the other vertices appear only
perturbatively).  
For two-particle scattering, this infinite set of diagrams is easily summed
as a geometric series.  
In three-body systems, large scattering lengths necessitate three-body
input at leading order to remove regularization dependence
\cite{Bedaque:1998kg,Bedaque:1999ve}.
The shallowness of the deuteron and triton bound states ensures that
the physics of large scattering lengths is dominant in these systems.%

At finite density, the class of diagrams with $C_0$ vertices is much
larger
than in free space, since it also includes tadpoles, particle-hole
rings, hole-hole rings, and so on.
These diagrams are not simply summed and    
have not been calculated numerically.
As an alternative to a direct numerical solution, 
we can seek additional expansions.
Possibilities include geometric \cite{Steele:2000qt}, 
strong-coupling, or large-$N$ expansions.  
In the present work, we investigate the most immediate extension that
is nonperturbative in $\kf a_s$ by adopting an 
effective action formalism, which is a natural framework for implementing
nonperturbative resummations.
As we illustrate below, the loop expansion in the auxiliary field
formulation for spin-independent forces is equivalent to a 
systematic $1/N$ expansion,
where the relevant ``$N$'' is the spin-isospin degeneracy $g$.

In Coleman's classic lecture on the $1/N$ expansion, he identifies two
reasons to pursue such expansions \cite{COLEMAN88}.  
First, they can be used to analyze
model field theories so that intuition beyond perturbation theory can
be developed.  
They provide  nontrivial, but tractable  
examples  (at least qualitatively)
to build intuition about phenomena such as asymptotic freedom,
dynamical symmetry breaking, dimensional transmutation, and
nonperturbative confinement.  We seek similar controlled insight into 
nonperturbative effects in nonrelativistic many-body
systems. 
The large-$g$ expansion has a well-defined power counting 
that sums certain classes of diagrams to all orders
to provide a systematic expansion
for systems where $\kf a_s$ is small (natural scattering length)
but $g\kf a_s/\pi$ is order unity or greater.
The  expansion for natural scattering length  has the
interesting feature that it is nonperturbative in the medium while the
description of free-space scattering is perturbative. 
For large $a_s$, the expansion may not be fully systematic, but 
could still provide insight into the nonperturbative many-body physics. 

The second reason for pursuing $1/N$ expansions is that they may be
relevant to physical systems of interest.  Coleman (and many others)
considered quantum chromodynamics (QCD) with a large number $N_c$ of
colors, arguing that the large $N_c$ world is in many ways close to
the real world of $N_c=3$.  
But while $1/N$ expansions of QCD and other relativistic field
theories are common \cite{Brezin:eb}, 
they have been exploited much less widely in nonrelativistic
many-fermion physics and very little in applications to nuclear systems.
In fact, the principle applications have been to
one-dimensional Fermi systems \cite{Calog75,Yoon77}.
However, there have been some applications of $1/N$ expansions 
in relativistic approaches to nuclear systems. 
In Ref.~\cite{TANAKA93}, e.g., the Walecka model was
studied in a $1/N$ expansion by extending the $SU(2)$ isospin 
symmetry to $SU(N)$.
To our knowledge, the present work is the first on effective action EFT 
and the $1/N$ expansion for nonrelativistic Fermi systems
with short-range interactions in three dimensions.

In Ref.~\cite{NEGELE88}, expansions in $1/g$ are discussed, but with
the caveat that they are not likely relevant to nuclear systems, even
though $g=4$ would seem to be large enough.  
This is based on the observation that with empirical nonrelativistic
interactions, exchange (Fock) contributions are larger than direct
(Hartree) contributions, whereas the former should be suppressed by $1/g$
relative to the latter.  However, this dismissal may be premature.  In
a different phenomenological representation of the problem, using covariant
interactions, the Hartree pieces do, in fact, dominate 
\cite{Furnstahl:1999ff}.  Furthermore,
the expansion may be useful to describe part of the physics, such as
if the long-distance pion physics is removed~\cite{Steele:1998zc}
or averaged out \cite{Furnstahl:2001un}.
Indeed, analyses of 
phenomenological energy functionals fit to bulk nuclear
properties  suggest a robust power counting with short-distance scale
$\Lambda$ roughly 600\,MeV \cite{Furnstahl:2000in}.  At nuclear matter
equilibrium densities, this would imply $\kf/\Lambda$ is less than one, but
$g\kf/\Lambda$ is greater than one! 
(However, the expansion parameter may actually be $g\kf/\Lambda\pi$,
as suggested by the analysis in Sect.~\ref{sec:results}.)

An interesting
special case of the $g \rightarrow \infty$ 
limit follows if we also take the
Fermi momentum $\kf$ to zero, with the density (which is proportional
to $g\kf^3$) held constant.
We refer to this procedure
as the ``Bose limit,'' because it generates the
ground state of a dilute Bose system
(under the assumption that the ground state evolves adiabatically from
the noninteracting state).
This limit was noted long ago in Refs.~\cite{GENTILE40} and
\cite{SHUBERT46} but was not exploited until Brandow used it in
comparing fermionic ($^3$He) and bosonic ($^4$He) many-body
systems~\cite{BRANDOW71}.
In Ref.~\cite{JACKSON94}, Jackson and Wettig used the Bose limit   
in analyzing minimal many-body approximations and cleanly
derived the
leading corrections to the Bose energy  
(see also Ref.~\cite{NEGELE88} 
in this connection.)
     
In Sect.~\ref{sec:auxiliary}, 
we summarize the effective action formalism for fermions
using an auxiliary field.  In Sect.~\ref{sec:dilute}, 
we review the EFT treatment
of a dilute Fermi system with short-range, spin-independent 
interactions and then
apply the effective action formalism.  
We carry out the loop expansion, which is seen to correspond to a
$1/g$ expansion.
In Sect.~\ref{sec:results},
we calculate the energy per particle explicitly to
next-to-leading order (NLO) and perform a stability
analysis of the ground state.
The Bose limit is considered in Sect.~\ref{sec:bose}, 
and Sect.~\ref{sec:summary} contains
a summary of our results and conclusions, along with plans for
further investigations.

\section{Effective action for fermions with auxiliary fields}
\label{sec:auxiliary}

In this section, we review the effective action formalism for
fermions with auxiliary fields at zero temperature ($T=0$). 
We adopt the notation and spirit of the general discussion of
effective actions in Ref.~\cite{PESKIN95}  
and largely follow the specific treatment of nonrelativistic fermions
by Fukuda et al.~\cite{FUKUDA94}. 
Our discussion will be somewhat 
schematic in order to focus on the new aspects of the EFT treatment. 
Details of the renormalization of the effective action
and caveats related to convergence of the path integrals, 
Wick rotations from Euclidean space, zero-temperature
limits, and so on are well documented and can be 
found in the standard literature 
\cite{COLEMAN88,NEGELE88,PESKIN95,FUKUDA94,ITZYKSON80,WEINBERG96}.

Consider a system of fermions
in an external potential $v(x)$ interacting via a spin-independent, 
local two-body interaction $U_0(x-y)$.
(The extension to many-body forces in the context of an EFT
is outlined in Appendix \ref{sec:appho}.)  
Such a system is described by the Lagrangian
\bea
{\cal L} &=& \psidal(x) \left( \idt +\frac{\nabsq}{2M}+\mu+v(x)\right) 
    \psi_\alpha(x) -\frac{1}{2} \psidal(x) \psidbe(y) U_0(x-y) 
    \psi_\beta(y) \psi_\alpha(x)\,,
\eea
where $\alpha,\, \beta$ are spin indices,
$\mu$ is the chemical potential, and $M$ the fermion mass.
We define a generating functional $Z[J]$ and
an energy functional $E[J]$ via the relation
(we follow the notation from Ref.~\cite{PESKIN95})
\beq
Z[J]=e^{-iE[J]}=\int{\cal D}\psid{\cal D}\psi\,
e^{i\int\! d^4 x \,[{\cal L} +J(x)\psidal(x)\psi_\alpha(x)]}\,,
  \label{eq:Zorig}
\eeq
where $J(x)$ is an external source.%
\footnote{Below we will use a generating functional with an external
source $J$ coupled to an auxiliary field rather than to
$\psi^\dagger\psi$.  These functionals generate the same observables;
see Ref.~\cite{FUKUDA94} for a discussion of the relationship between
them.}
For simplicity, normalization factors are considered to be implicit
in the functional integration measure.

The strategy is first to reduce $Z[J]$ to a Gaussian integral in the
fermion Grassmann fields by ``integrating in'' an 
auxiliary field \cite{HandS,NEGELE88}. 
Using the identity \cite{FUKUDA94}
\beq
1=\frac{\int {\cal D}\sigma\exp\left(\frac{i}{2}\int\! d^4 x \int\! d^4 y\,
  [\sigma(x)-\psidal(x)\psi_\alpha(x)]U_0(x-y)[\sigma(y)
   -\psidbe(y)\psi_\beta(y)]\right)}
  {\int {\cal D}\sigma\exp\left(\frac{i}{2}\int\! d^4 x \int\! d^4 y\,
   \sigma(x) U_0(x-y) \sigma(y)\right)}\,,
\label{eq:pathid}
\eeq
we introduce an auxiliary field $\sigma$ with bosonic quantum
numbers and obtain for the generating functional
with $J=0$ (i.e., the partition function)
\bea
Z[J]|_{J=0} &=& \int {\cal D}\,\psid{\cal D}\psi\, {\cal D}\sigma\exp\bigg[
   i\int\!d^4x\,
  \psidal(x) \bigg( \idt +\frac{\nabsq}{2M}+\mu+v(x)\bigg)  \psi_\alpha(x)
\nonumber \\
 && \null  
 -\psidal(x)\psi_\alpha(x)  \int\! d^4 z\, U_0(x-z) \sigma(z)
 +\frac{1}{2} \sigma(x) \int\! d^4 y\, U_0(x-y) \sigma(y)
 \bigg]  \,,
\eea
where 
the denominator of Eq.~(\ref{eq:pathid}) has been absorbed
into the normalization of $Z$.
The ${\cal D}\psid {\cal D}\psi$ path integral is now Gaussian
and leads to a determinant of the operator between $\psid$ and $\psi$,
which we identify as an inverse fermion propagator:
\bea
  G^{-1}(x,y)\,\delta_{\alpha\beta} &\equiv& 
   \biggl[ \idt +\frac{\nabsqx}{2M}+\mu+v(x)-\int\! d^4 z U_0(x-z) \sigma(z)
   \biggr] \delta^4 (x-y)\,\delta_{\alpha\beta} \ .
\eea
Note that this propagator still depends on the field $\sigma$
but is manifestly diagonal in the spin (or flavor) indices.
We exponentiate the determinant using $\det A = \exp[\Tr\ln A]$
and reintroduce a source term,
now coupled to the $\sigma$ field, 
to obtain
\bea
  Z[J]&=&e^{-iE[J]}= \int {\cal D}\sigma \exp\bigg(g\Tr\ln\bigg[
   G^{-1}(x,y) 
    \bigg] \bigg) \nonumber\\
  &&  \null \times
    \exp\bigg(\frac{i}{2} \int\! d^4 x \!\int\! d^4 y\, \sigma(x) U_0(x-y) 
    \sigma(y)\bigg) \exp\bigg(i\int\! d^4 x J(x)\sigma(x)\bigg)\,,
  \label{eq:ef1}
\eea
where the spin/flavor trace has been performed in the
first term, leading to  the spin/flavor degeneracy 
factor $g$ (e.g., for electrons or neutrons, $g=2$ and for symmetric
nuclear matter, $g=4$).
The remaining trace in the first line of Eq.~(\ref{eq:ef1})
is over space-time.

We define the \lq\lq classical field'' $\sigma_c(x)$ 
in the presence of $J(x)$ by the ground state expectation value
of $\sigma(x)$:
\beq
  \sigma_c(x) \equiv
  \langle \Omega |\sigma(x)|\Omega\rangle_J 
  = -\frac{\delta}{\delta J(x)}E[J]
  = -i\frac{\delta}{\delta J(x)}\ln Z[J]
    \,,
\eeq
and define the effective action
\beq
  \Gamma[\sigma_c]\equiv -E[J]-\int\! d^4 x\, J(x) \sigma_c(x)\,,
\eeq
in the usual way \cite{PESKIN95}. The effective action has the property
\beq
\frac{\delta}{\delta \sigma_c(x)} \Gamma[\sigma_c]=-J(x)\,,
\eeq
and  the solutions for $J(x)=0$ represent the stable
quantum states \cite{PESKIN95}. 
At the minimum $\sigma_c^0$ (with $J(x)=0$) of a \emph{uniform}
system, the energy density $\edens$ of
the ground state is related to the effective action by
\beq
\left.\Gamma[\sigma_c^0]\right|_{J=0}= -VT \edens \,,
\eeq
where $VT$ is the space-time volume of the system.
More generally, at finite density
we must examine spatially dependent $\sigma_c^0$ to
find the absolute ground state.

To evaluate $\Gamma[\sigma_c]$ in a loop expansion,
we write $\sigma=\sigma_c+\eta$ and expand Eq.~(\ref{eq:ef1})
in quantum fluctuations $\eta$ around the classical field $\sigma_c$.
Once again we seek a Gaussian path integral, this time in terms of $\eta$,
while using a source coupled to $\eta$ to treat the residual $\eta$-dependent
terms perturbatively (by removing them from the path integral in favor
of functional derivatives with respect to the source).
The expansion, integration, and subsequent Legendre transformation
are standard and we simply quote the result for the
effective action from 
Refs.~\cite{FUKUDA94,PESKIN95}:
\bea
\Gamma[\sigma_c]&=&\frac{g}{i}\,\Tr\ln[G_H^{-1}(x,y)]+\frac{1}{2}
 \int\! d^4 x \int\! d^4 y \,\sigma_c(x) U_0(x-y) \sigma_c(y)
\nonumber\\
&&+ \frac{i}{2}\Tr\ln\left[D_\sigma^{-1}(x,y)\right]
   + \int\! d^4 x \,\delta\Left[\sigma_c]+ 
     \mbox{(connected 1PI-diagrams)}\,.
\label{eq:effact}
\eea
In Eq.~(\ref{eq:effact}), we have introduced the inverse Hartree
propagator (which depends on $\sigma_c$ but not $\eta$)  
\bea
  G_H^{-1}(x,y) &\equiv& 
   \biggl[ \idt +\frac{\nabsqx}{2M}+\mu+v(x)-\int\! d^4 z U_0(x-z) \sigma_c(z)
   \biggr] \delta^4 (x-y) \ ,
\eea
and
the inverse $\sigma$ propagator
\beq
D_\sigma^{-1} (x,y)\equiv 
  -iU_0(x-y) +
  g\int\! d^4 z_1 \int\! d^4 z_2\,
   G_H (z_1,z_2) U_0(z_2-x) G_H (z_2,z_1) U_0(z_1-y)
   \,,
\eeq
which originates with the part of the exponential in
$Z[J]$ that is quadratic in $\eta$
after expanding.
The counterterm Lagrangian $\delta{\cal L}$ has been included in
Eq.~(\ref{eq:effact}) for completeness; 
however, with the regularization/renormalization
procedure applied here (DR/MS), we will not need it 
explicitly \cite{PESKIN95}.

The two  propagators are obtained by solving the
 equations
\bea
\int\! d^4 z\, G_H^{-1}(x,z)G_H(z,y)&=&\delta^4 (x-y)\,,\nonumber\\
\int\! d^4 z\, D_\sigma^{-1}(x,z)D_\sigma(z,y)&=&\delta^4 (x-y)\,,
\eea
with  appropriate boundary conditions for $G_H$ and $D_\sigma$ 
(discussed below).
The \lq\lq connected 1PI-diagrams'' in Eq.~(\ref{eq:effact})
are built from the propagator $D_\sigma$ for the $\sigma$ field
and the vertices
\beq
V_m(x_1,..,x_m)=\int\! d^4 y_1...\int\! d^4 y_m\,
 G_H(y_m,y_1)\,U_0(y_1-x_1).. G_H(y_{m-1},y_m)\,
 U_0(y_m-x_m)\,,
\eeq
where $m=3,\ldots,\infty$. 
We follow Ref.~\cite{PESKIN95}
and use the freedom of a counterterm $\delta J$ to
ensure that $\sigma_c$ is the same to all orders 
in the expansion (i.e., $\langle \eta \rangle = 0$), 
which simplifies the bookkeeping.
As a consequence, all diagrams that are
1-particle reducible with respect to the $\sigma$ propagator are
cancelled and only 1-particle irreducible diagrams
have to be included in calculations of the effective action.
In the next section, we show how this expansion, when applied to the
lowest-order EFT potential, is an expansion in inverse powers of
$g$.

\section{Dilute Fermi systems with short-range interactions}
\label{sec:dilute}

In this section, we apply the effective action formalism to
the Fermi gas with short-range, spin-independent interactions.
We  calculate the effective action 
to one loop explicitly
and demonstrate that the loop expansion of the effective
action corresponds to a large $N$ expansion. 

\subsection{Short-range EFT and Low-density Expansion}
We consider
a general local Lagrangian for a nonrelativistic fermion
field that is invariant under Galilean, parity, and time-reversal
transformations:
\bea
  {\cal L}  &=&
       \psi^\dagger \biggl[i\partial_t + \frac{\nab^{\,2}}{2M}\biggr]
                 \psi - \frac{C_0}{2}(\psi^\dagger \psi)^2
            + \frac{C_2}{16}\Bigl[ (\psi\psi)^\dagger
                                  (\psi\galnab^2\psi)+\mbox{ h.c.}
                             \Bigr]
  \nonumber \\[5pt]
   & & \null +
         \frac{C_2'}{8} (\psi \galnab \psi)^\dagger \cdot
             (\psi\galnab \psi)
%         - \frac{D_0}{6}(\psi^\dagger \psi)^3 
+  \ldots\,,
  \label{eq:lag}                                                   
\eea
where $\galnab=\overleftarrow{\nabla}-\nab$ is the Galilean invariant
derivative and h.c.\ denotes the Hermitian conjugate.
The terms proportional to $C_2$ and $C_2'$ contribute to $s$-wave and
$p$-wave scattering, respectively, 
while the dots represent terms with more derivatives and/or more
fields.
The Lagrangian Eq.~(\ref{eq:lag}) represents a particular 
canonical form, which can be reached via field redefinitions.
For example,
higher-order terms with time derivatives are omitted, as they can be
eliminated in favor of terms with spatial derivatives.                        

To reproduce the results in Ref.~\cite{HAMMER00},
we can write a generating functional with the lagrangian
Eq.~(\ref{eq:lag}) and Grassmann sources coupled to $\psi^\dagger$ and
$\psi$, respectively.
Perturbative expansions for Green's functions follow by taking
successive functional derivatives, and the ground state energy density
follows by applying the linked cluster theorem (see Ref.~\cite{NEGELE88}). 

\begin{figure}[t]
\centerline{\includegraphics*[width=14cm,angle=0]{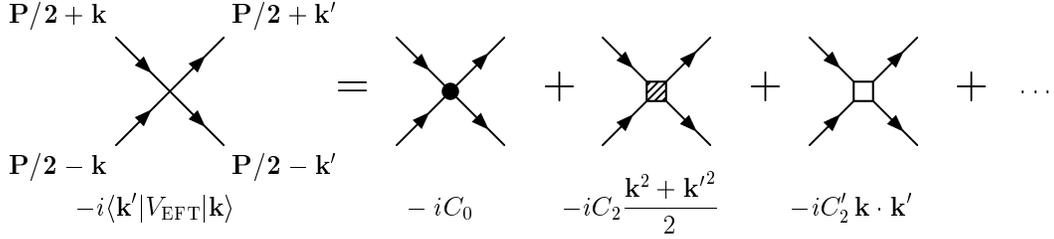}}
\vspace*{-.0in}
\caption{Feynman rules for $\langle\vec{k}'|V_{\rm EFT}|\vec{k}\rangle$.
$\vec{P}$ is the total momentum in the center of mass.
The spin indices have been suppressed.}
\label{fig:eftvertex}
\end{figure}

The scattering amplitude for fermions in the vacuum is simply related
to a sum of Feynman graphs computed according to this lagrangian.
The terms in Eq.~(\ref{eq:lag})
involving only four fermion fields reduce in momentum
space to simple polynomial vertices
that are equivalent to a momentum expansion of an EFT potential for
particle-particle scattering,
\beq
  \langle \vec{k'} | V_{EFT} | \vec{k}\rangle
      = C_0 + C_2 (\vec{k'}^2 +\vec{k}^2)/2
         + C_2' \,\vec{k'}\cdot\vec{k}
         + \ldots \ ,
  \label{eq:effpot}
\eeq
(see Fig.~\ref{fig:eftvertex} for the Feynman rules).
Because of Galilean invariance, the interaction depends only on the
relative momenta  $\vec{k}$ and $\vec{k'}$ of the incoming and outgoing
particles.                                
The coefficients $C_0$, $C_2$, and $C_2'$ can
be obtained from matching the EFT
to a more fundamental theory or to
(at least) three independent pieces of experimental data.
It is important to note that Eq.~(\ref{eq:effpot}) is {\em not\/}
simply the term-by-term momentum-space expansion of an underlying potential
because the coefficients also contain short-distance contributions
from loop graphs. 
By matching to the effective range expansion,
we can express the $C_{2i}$ in terms of the effective range parameters:
\beq
      C_0 = \frac{4\pi a_s}{M},\qquad C_2=C_0 \frac{a_s r_s}{2},
               \quad \mbox{and}\quad
          C_2' = \frac{4\pi a_p^3}{M}
      \label{C2imatch}    \,,
\eeq
where $a_s$ and $r_s$ are the $s$-wave  scattering length and
effective range, respectively, and $a_p$ is the $p$-wave scattering
length.
(The implied renormalization prescription here is DR/MS, as described
in Ref.~\cite{HAMMER00}.)

\begin{figure}[t]
\centerline{\includegraphics*[width=14cm,angle=0]{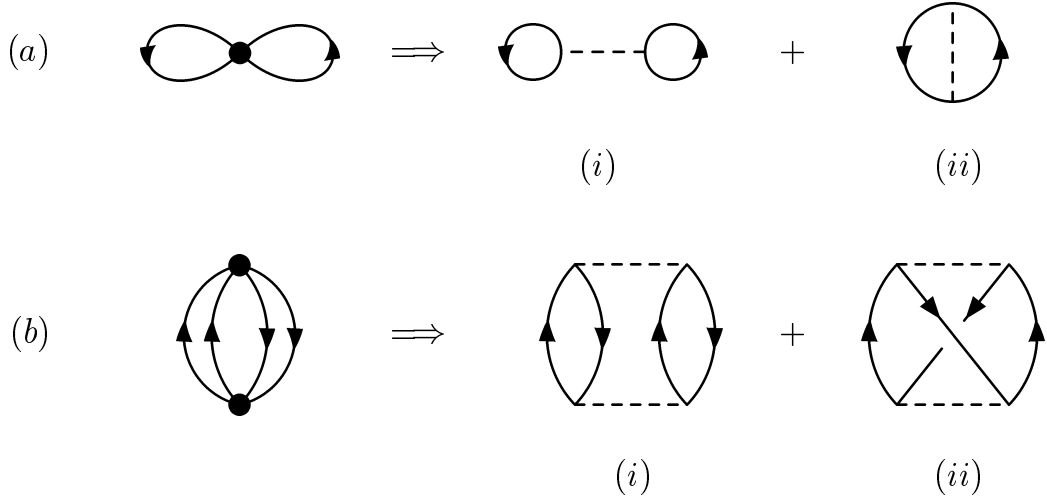}}
\vspace*{-.0in}
\caption{Hugenholtz diagrams contributing to the energy density of the 
dilute Fermi gas at ${\cal O}(\kf^6)$ (a) and ${\cal O}(\kf^7)$ (b).
The right-hand side shows the two possible spin-contractions for
each of the two diagrams.}
\label{fig:edlonlo}
\end{figure}

The energy density at finite density
can be calculated from diagrams with no external
lines using the  vertices in Fig.~\ref{fig:eftvertex} and the propagator
\beq
    G_0 (k_0,\vec{k})_{\alpha\beta}=\delta_{\alpha\beta}
    \left( \frac{\theta(k-\kf)}{k_0-\omega_\vec{k}+i\epsilon}
      +\frac{\theta(\kf-k)}{k_0-\omega_\vec{k}-i\epsilon}\right) \,,
      \label{eq:freeprop_fd}
\eeq
where $\omega_\vec{k}=\vec{k}^2/(2M)$.
In this approach, which is applicable to a dilute Fermi system at $T=0$, 
the finite density
boundary conditions are imposed by hand in Eq.~(\ref{eq:freeprop_fd}),
rather than using a chemical potential and 
inverting it at the end \cite{FETTER71}.
The leading contribution to the energy density
is the kinetic energy of the noninteracting Fermi gas
\beq
{\cal E}_{kin}=g\int\frac{d^3 k}{(2\pi)^3}\, \omega_\vec{k}\, \theta(\kf-k)
=n\, \frac{3}{5}\, \frac{\kf^2}{2M}\,.
\eeq
where 
\beq
        n=g\kf^3/(6\pi^2)
        \label{eq:density} 
\eeq
is the number density of the system.
In the low-density limit, the leading corrections come from the diagrams 
shown on the left-hand side of  Fig.~\ref{fig:edlonlo}.

For tadpole diagrams such as in Fig.~\ref{fig:edlonlo}(a), 
a convergence factor $\exp(ik_0\eta)$ must be included, with the limit 
$\eta\to 0^+$ being taken after the contour integrals have been 
carried out. This procedure automatically takes into account that such 
lines must be hole lines. Diagram~\ref{fig:edlonlo}(b) 
does not need the convergence factor,
but has a power law UV divergence and must be renormalized. 
A convenient regulator is dimensional regularization with minimal 
subtraction (DR/MS) \cite{HAMMER00}. 
For a detailed description of the Feynman rules and the 
calculation of the energy density for a dilute Fermi gas
to ${\cal O}(\kf^9\ln\kf)$, including three-body contributions, 
see Ref.~\cite{HAMMER00}.
To ${\cal O}(\kf^7)$ the energy density 
is \cite{FETTER71}\footnote{Note that the low-density expansion 
corresponds to an expansion in powers of $\kf$ and that the overall factor 
of density implicitly contains three powers of $\kf$.}
\bea
        {\cal E} &=& n
        \frac{\kf^2}{2M}
        \biggl[ \frac{3}{5} + (g-1)\biggl\{ \frac{2}{3\pi}(\kf a_s)
          + \frac{4}{35\pi^2}(11-2\ln 2)(\kf a_s)^2 \bigg\}\bigg]
     \label{eq:edkf7}
\eea                                        
where the first term is the kinetic energy of the noninteracting
Fermi gas and the second and third terms are corrections 
from the diagrams in Fig.~\ref{fig:edlonlo}(a) and (b), respectively.

In the low-density expansion of Eq.~(\ref{eq:edkf7}), it is assumed that
$g \sim {\cal O}(1)$, and therefore the direct and exchange contributions
to Fig.~\ref{fig:edlonlo}(a) are counted at the same order in the
$\kf$ expansion (and similarly with
higher-order contributions).  Thus, Hugenholtz diagrams, which combine these
contributions, are particularly efficient for this expansion.
If we study systems where $g$ can be considered large, then we will
need a new power counting that 
assigns direct and exchange contributions to different orders in the
EFT expansion.
The isolation of different $g$ dependencies 
is accomplished in the diagrams on the right-hand side of
Fig.~\ref{fig:edlonlo}, which follow from the introduction of an auxiliary
field.  We consider a consistent power counting and summation
of such diagrams in the next section.

\subsection{Effective Action for Short-Range Interactions}

Here we will calculate the energy density for the Fermi gas 
with short-range interactions using
the effective action formalism. The loop expansion of the 
effective action does not correspond to 
a simple low-density expansion but resums an infinite class of diagrams.
We will only explicitly
consider the momentum independent $C_0$ interaction
from Eq.~(\ref{eq:lag}). The other $C_{2i}$ vertices and many-body
forces can be included perturbatively (see Appendix \ref{sec:appho}).

\begin{figure}[t]
\centerline{\includegraphics*[width=10cm,angle=0]{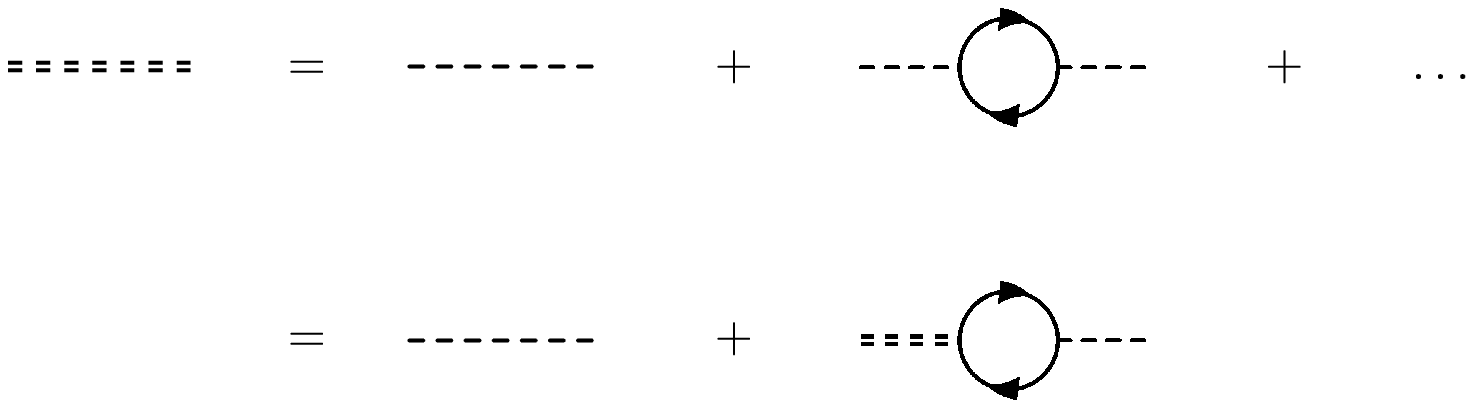}}
\vspace*{-.0in}
\caption{Integral equation satisfied by the $\sigma$ propagator 
(double dashed line). The single dashed line indicates the  
bare $\sigma$ (a single potential exchange) 
while the solid lines are Hartree propagators.}
\label{fig_sigprop}
\end{figure}

The interaction $U_0$ from the previous section
can then be specified as
\beq
U_0 (x-y)= C_0 \delta^4 (x-y)\,,
\label{eq:u0c0}
\eeq
which corresponds to the second term in Eq.~(\ref{eq:lag}).%
\footnote{Strictly speaking there is a contribution proportional
to $\psi^\dagger\psi$ from normal ordering, but this can be absorbed into a
counterterm for the chemical potential.}
Inserting Eq.~(\ref{eq:u0c0}) into Eq.~(\ref{eq:effact}), we obtain
\bea
\Gamma[\sigma_c]&=&\frac{g}{i}\,\Tr\ln[G_H^{-1}(x,y)]+\frac{C_0}{2}
 \int\!d^4x\,\sigma_c(x)^2 \nonumber\\
& & 
 +\frac{i}{2}
 \Tr\ln[D_\sigma^{-1}(x,y)]
  \nonumber\\
& &+ \mbox{ (connected 1PI-diagrams)}\,.
\label{eq:effactapp}
\eea
The inverse Hartree and $\sigma$ propagators simplify to
\bea
  G_H^{-1}(x,y) &\equiv& 
   \left[ \idt +\frac{\nabsqx}{2M}+\mu+v(x)-C_0 \sigma_c(x)
   \right]  
   \delta^4 (x-y)
   \ ,
  \label{eq:hpropc0}
  \\
  D_\sigma^{-1} (x,y)&\equiv& 
  \null -iC_0\delta^4(x-y) 
  +
  g C_0^2\,
  G_H (y,x) G_H (x,y)
  \,.
  \label{eq:spropc0}
\eea
The $\sigma$ propagator $D_\sigma(x,y)$ is given by the sum of 
all fermion ring diagrams and satisfies the integral equation
\beq
  D_\sigma(x,y) = \frac{i}{C_0}\delta^4 (x-y)
     - i g C_0 \int\! d^4z \, D_\sigma(x,z) G_H(z,y) G_H(y,z)    
   \,, 
\label{eq:dsigmaeq}
\eeq
which is illustrated in Fig.~\ref{fig_sigprop}.
The solid lines denote Hartree propagators $G_H(x,y)$
while the double dashed line denotes the $\sigma$ propagator 
$D_\sigma(x,y)$. 
The single-dashed lines represent the bare $\sigma$ propagator 
given by the inhomogeneous term in Eq.~(\ref{eq:dsigmaeq});
it corresponds a single $\sigma$ exchange.  
The bare $\sigma$ propagator carries a factor 
$1/C_0$ while each coupling of the bare $\sigma$ to fermions carries a
factor $C_0$, such that a four-fermion interaction is proportional
to $C_0$ as required.

\begin{figure}[t]
\centerline{\includegraphics*[width=10cm,angle=0]{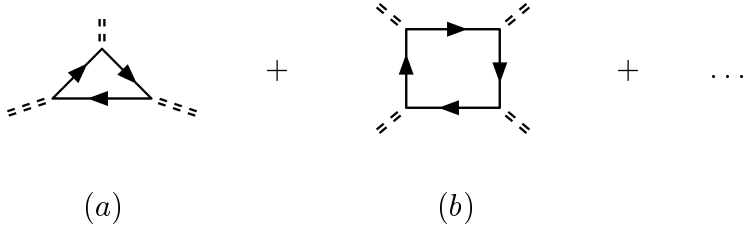}}
\vspace*{-.0in}
\caption{Vertices (a) $V_3$  and (b) $V_4$
for the 1PI-diagrams from Eq.~(\ref{eq:etavertc0}).
The solid lines indicate Hartree propagators $G_H$ while the double dashed 
lines indicate the $\sigma$ propagator $D_\sigma$.}
\label{fig_sigvert}
\end{figure}

\begin{figure}[t]
\centerline{\includegraphics*[width=10cm,angle=0]{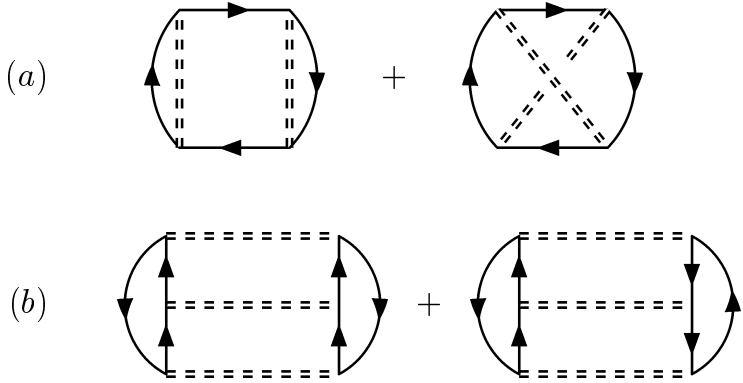}}
\vspace*{-.0in}
\caption{\lq\lq Connected 1PI-diagrams'' contributing to the effective 
action at two loops.
The solid lines indicate Hartree propagators $G_H$ while the double dashed 
lines indicate the $\sigma$ propagator $D_\sigma$.}
\label{fig_dian2lo}
\end{figure}

The \lq\lq connected 1PI-diagrams'' are built from $D_\sigma(x,y)$
and the vertices
\beq
V_m(x_1,..,x_m)=(C_0)^m
 G_H(x_m,x_1)\ldots G_H(x_{m-1},x_m)\,,
\label{eq:etavertc0}
\eeq
where $m=3,\ldots,\infty$. 
In Fig.~\ref{fig_sigvert}, we show
the lowest-$m$ vertices, $V_3$  and $V_4$,  as an example.
The solid lines indicate Hartree propagators $G_H$ while the double dashed 
line indicate the $\sigma$ propagator $D_\sigma$.
The \lq\lq connected 1PI-diagrams'' can be arranged in
a consistent loop expansion, where the corresponding expansion parameter is the
total number of loops minus the number of fermion loops
[see Eq.~(\ref{eq:nu2})]. 
The leading corrections to the terms
explicitly given in Eq.(\ref{eq:effactapp}) start at two loops.
As an example, we show the \lq\lq connected 1PI-diagrams'' at two loops
in  Fig.~\ref{fig_dian2lo}.
All other diagrams have at least three  loops.

The status of the loop expansion as an expansion in $1/g$ can be seen
by returning to the generating functional of Eq.~(\ref{eq:ef1}), but
now specialized to the contact interaction of Eq.~(\ref{eq:u0c0}):
\bea
  Z[J] &=& e^{-iE[J]}= \int {\cal D}\sigma 
  \,\exp\left\{g\,\Tr\ln \left[
   \idt + \frac{\nabsq}{2M}+\mu - C_0\sigma(x) \right]\right\}  \nonumber\\
  &&  \qquad\qquad\qquad\qquad \null  
  \times \exp\left\{i \int d^4 x \, \left(\frac{1}{2} C_0\sigma(x)^2 
  + J(x)\sigma(x)\right)\right\} \ .
  \label{eq:genfun}
\eea
If we scale $C_0$ and $\sigma$ according to
$C_0 = c_0/g$ and $\sigma = g \sigma'$, then we find that
all of the $g$ dependence reduces to a single overall factor
multiplying the path integral exponent. 
Thus, a saddle
point evaluation of the integral is just the loop expansion with the
loop counting parameter equal to $1/g$ (which plays the same role
as $\hbar$ in the usual discussion) \cite{NEGELE88}.  Below we will
examine the contributions order by order.

\subsection{Leading-order Effective Potential}

In the following we carry out the effective action formalism for
a uniform Fermi gas with short-range interactions.
For simplicity,
we work without an explicit chemical potential and 
incorporate the appropriate finite-density
boundary conditions in the propagator
\cite{FETTER71,HAMMER00}. 
We also set the external potential $v(x)$ to 
zero, since we are interested at present in an infinite system.
We assume that the ground state is uniform.
As a consequence, the classical field $\sigma_c(x)$ is independent
of $x$ and we can set $\sigma_c(x)=\sigma_c$ in Eq.~(\ref{eq:hpropc0}).
We reassess this assumption in Sect.~\ref{sec:results}.

The first step is to calculate the contribution to the
effective potential given by the first line in 
Eq.~(\ref{eq:effactapp}). We refer to this contribution as the tree-level
contribution since it does not depend on $D_\sigma$.
The $\Tr\ln D_\sigma^{-1}$ in the second line of 
Eq.~(\ref{eq:effactapp}) gives the one-loop part of the effective
potential. Since $G_H^{-1}$ is diagonal in momentum space, the $\Tr\ln$ 
can be evaluated as
\beq
\Tr\ln G_H^{-1}= VT \int\frac{d^3 p}{(2\pi)^3}\int\frac{dp_0}{2\pi}
\ln(p_0-e_\vec{p})
\eeq
where 
\beq
    e_\vec{p}=\vec{p}^2/(2M)+C_0 \sigma_c
\eeq
and $VT$ is the
spacetime volume.
The $dp_0$ integral can be regularized by taking a derivative
with respect to $e_\vec{p}$ and integrating back after the 
$dp_0$ integral is carried out. 
(The lower integration limit can be taken to be independent
of $\vec{p}$ and will not contribute to the $d^3p$ integration in
dimensional regularization.)
The resulting integral has the form
of a tadpole diagram with a modified propagator. 
The appropriate boundary
conditions can be included by the replacement
\beq
\frac{1}{p_0-e_\vec{p}}\quad\Longrightarrow\quad
G_H(p)=\frac{\theta(p-\kf)}{p_0
-e_\vec{p} +i\epsilon}+\frac{\theta(\kf-p)}{p_0
-e_\vec{p}-i\epsilon}\,.
\label{eq:GHC0}
\eeq
Applying the convergence factor $\exp(i\eta p_0)$ and taking the 
limit $\eta\to 0^+$ after the contour integral has been carried out,
we obtain
\beq
\frac{g}{i}\Tr\ln G_H^{-1} = -g VT \int\frac{d^3 p}{(2\pi)^3}
\left(\frac{\vec{p}^2}{2M} +C_0 \sigma_c \right)\theta(\kf-p)\,.
\eeq
The $d^3 p$ integral is now immediate and the full tree
contribution to the effective potential from the first line of 
Eq.~(\ref{eq:effactapp}) is
\beq
\Gamma_{LO} [\sigma_c]=VT\left(\frac{C_0}{2}\sigma_c^2-\frac{3}{5}
\frac{\kf^2}{2M}n-C_0\sigma_c\, n\right)\,.
\label{eq:effactapp0}
\eeq

By requiring the effective action to be stationary,
\beq
  \left.\frac{\delta\Gamma}{\delta\sigma_c}\right|_{\sigma_c=\sigma_c^0}
  = 0
  \quad \Longrightarrow\quad \sigma_c^0=n\,,
\eeq
and substituting $\sigma_c^0$ back into Eq.~(\ref{eq:effactapp0}),
we obtain the energy density at tree level 
\beq
{\cal E}_{LO} = n\left( \frac{3}{5}\frac{\kf^2}{2M}+\frac{C_0}{2}n\right)
= n\frac{\kf^2}{2M}\left( \frac{3}{5}+g\frac{2}{3\pi}\kf a_s\right)\,.
\label{eq:eden0}
\eeq
Note that this expression includes the Hartree term [diagram
$(i)$ in Fig.~\ref{fig:edlonlo}(a)] while the Fock
term [diagram $(ii)$ in Fig.~\ref{fig:edlonlo}(a)]
will appear in the $\Tr\ln$ of the inverse $\sigma$ propagator
in the second line of Eq.~(\ref{eq:effactapp}).
[Diagram $(i)$ of Fig.~\ref{fig:edlonlo}(b) will also appear
in the $\Tr\ln$ at one-loop order
while diagram $(ii)$ of Fig.~\ref{fig:edlonlo}(b)
will appear at two loops, as seen from the first equality
of Eq.~(\ref{eq:nu2}).] 
As a consequence, the loop expansion of the effective potential 
\emph{does not} 
correspond to the low-density expansion of the dilute Fermi gas where
the Hartree and Fock terms appear at the same order \cite{HAMMER00}. 

Instead, the loop expansion of the effective potential corresponds to
a $1/N$ expansion where $N$ is the spin degeneracy factor $g$.
This can be seen as follows:
We take the limit $g\to \infty$ for the energy per particle
$E/N={\cal E}/n$ by defining a new coupling 
\beq
c_0 = g\,C_0\,,
\label{eq:c0largeN}
\eeq 
and keep $c_0$ and $\kf$ fixed as $g \to \infty$. Using 
Eqs.~(\ref{eq:eden0},\ref{eq:c0largeN}), we find
\beq
\left.\frac{E}{N}\right|_{LO}=\frac{3}{5}\frac{\kf^2}{2M}+\frac{c_0}{2}
\frac{\kf^3}{6\pi^2}\,,
\eeq
which is ${\cal O}(1)$ in the $1/g$ expansion.
Note that Eq.~(\ref{eq:c0largeN}) implies that the scattering length
$a_s$ is ${\cal O}(1/g)$ in the large $g$ limit.  
Thus, scattering in free space
is perturbative, since a particle can scatter off only one other
particle, while the many-body problem is nonperturbative 
as $g \rightarrow \infty$ since a particle can scatter off $g$ other
particles.

\subsection{Next-to-leading Order Effective Potential}
We proceed to calculate the one-loop contribution to the 
effective potential given by  $\Tr\ln D_\sigma^{-1}$
in the second line of Eq.~(\ref{eq:effactapp}).
We will show below that this term is of ${\cal O}(1/g)$ and 
constitutes the first correction to Eq.~(\ref{eq:eden0}).
After performing a four-dimensional Fourier transform
\beq
G_H(x_1,x_2)=\int\frac{d^4 p}{(2\pi)^4}G_H(p) 
e^{-ip\cdot (x_1-x_2)}\,,
\eeq
the contribution from the second line in Eq.~(\ref{eq:effactapp})
to the effective potential becomes 
\bea
\Gamma_{NLO} [\sigma_c]&=&\frac{i}{2}\Tr_{x_1,x_2}\ln\left[
\int\frac{d^4 q}{(2\pi)^4} e^{-iq\cdot (x_1-x_2)} 
\left(  
  -i C_0 
  + g\, C_0^2\,
   \int\frac{d^4 p}{(2\pi)^4} G_H(p+q) G_H(p)
\right)
\right]\nonumber\\
&=&\frac{i}{2}VT \int\frac{d^4 q}{(2\pi)^4}\ln
\left( 
  -i C_0
  + g\, C_0^2\,
  \int\frac{d^4 p}{(2\pi)^4} G_H(p+q) G_H(p) 
\right)\,,
\label{eq:gam1}
\eea
where $G_H(p)$ is the Hartree propagator with the appropriate 
boundary conditions given in Eq.~(\ref{eq:GHC0}).

It is customary to define the polarization insertion \cite{FETTER71}
\beq
\Pi_0 (q)\equiv -ig \int\frac{d^4 p}{(2\pi)^4} G_0(p+q) G_0(p)=
-ig \int\frac{d^4 p}{(2\pi)^4} G_H(p+q) G_H(p)\,,
 \label{eq:polins}
\eeq
where $G_0(p)$ is given in Eq.~(\ref{eq:freeprop_fd}) and the
second equality is valid for a uniform system,
for which $\sigma_c$ is a constant that can be eliminated by a shift
in $p_0$.
By taking a derivative with respect to $C_0$ in Eq.~(\ref{eq:gam1})
and integrating back, we obtain 
\beq
\Gamma_{NLO} [\sigma_c]=-\frac{i}{2}VT \int_0^{C_0} \frac{dy}{y} 
\int\frac{d^4 q}{(2\pi)^4} \frac{y\,\Pi_0(q)}{
    1-y\,\Pi_0(q)}
\label{eq:effactapp1}
\eeq
where we have used that $\int\! d^4 q/(2\pi)^4\equiv 0$ in dimensional
regularization.
The first two terms in the expansion of $1/(1-y\,\Pi_0(q))$
in powers of $y\,\Pi_0(q)$
correspond to the Fock term (diagram $(ii)$ in Fig.~\ref{fig:edlonlo}(a))
and diagram $(i)$ in Fig.~\ref{fig:edlonlo}(b).
While all higher-order terms in the expansion  are 
UV finite and can be summed straightforwardly, those two
terms are special:
The diagrams in Fig.~\ref{fig:edlonlo}(a) are tadpoles that
require a convergence factor and the diagrams in 
Fig.~\ref{fig:edlonlo}(b) have  power law UV divergences
and need to be renormalized. Therefore, we 
subtract those two terms from Eq.~(\ref{eq:effactapp1})
and calculate them separately using the methods of
Ref.~\cite{HAMMER00}.

The remaining finite terms are easily summed by calculating
\beq
\widetilde{\Gamma}_{NLO} [\sigma_c]=-\frac{i}{2}VT \int_0^{C_0} \frac{dy}{y} 
\int\frac{d^4 q}{(2\pi)^4} \frac{(y\,\Pi_0(q))^3}{
    1-y\,\Pi_0(q)} \ .
\label{eq:effactapp1b}
\eeq
Performing the $dy$ integral in Eq.~(\ref{eq:effactapp1b}) leads to
\beq
\widetilde{\Gamma}_{NLO} [\sigma_c]=-\frac{i}{2}VT
\int\frac{d^4 q}{(2\pi)^4} \left[
C_0 \Pi_0(q)+\frac{1}{2} (C_0 \Pi_0(q))^2+\ln(1-C_0 \Pi_0(q)) \right]
\label{eq:effactapp1c}
\eeq
and it is straightforward to
verify that the first two terms in the integrand simply
cancel the first two terms in an expansion of the logarithm. 
Note the similarity of Eq.~(\ref{eq:effactapp1c}) to the expression
 [Eq.~(12.53)] for the correlation energy of a degenerate electron gas in
Ref.~\cite{FETTER71}.
Because
$\widetilde{\Gamma}_{NLO}$ does not depend on $\sigma_c$ (for uniform
background fields), it 
does not have to be minimized and is,
up to a factor of $-VT$, equal to the contribution to the energy 
density at ${\cal O}(1/g)$.
Since the energy density is real, we can write
\bea
\widetilde{\cal E}_{NLO} &=& \frac{\kf^5}{4\pi^3 M}
\int_0^\infty  v^2 dv  \int_0^\infty dv_0 \Bigg[
C_0 {\rm Im}\, \Pi_0(v_0,v)+C_0^2 {\rm Im}\, \Pi_0(v_0,v)
{\rm Re}\, \Pi_0(v_0,v) \nonumber\\
&& +\arctan \left(\frac{
C_0 {\rm Im}\,\Pi_0(v_0,v)}{C_0 {\rm Re}\,\Pi_0(v_0,v)-1}\right)\,,
\Bigg]
\label{eq:ed1}
\eea
where $v=|\vec{v}|$ and $(v_0,\vec{v})$ are dimensionless variables 
related to the four-momentum $(q_0,\vec{q})$ via
\beq
v=|\vec{q}|/\kf\qquad \mbox{and}\qquad v_0=Mq_0/\kf^2\,.
\eeq
Adding the contributions of the Fock term and diagram $(i)$
of Fig.~\ref{fig:edlonlo}(b),
the full contribution to the energy density is given by
\beq
{\cal E}_{NLO} =
        n \frac{\kf^2}{2M}
        \biggl[ -\frac{2}{3\pi}(\kf a_s)
          + g \frac{4}{35\pi^2}(11-2\ln 2)(\kf a_s)^2 \bigg]
         +  \widetilde{\cal E}_{NLO}          
          \,.
\label{eq:eden-1}
\eeq
Using Eqs.~(\ref{C2imatch}) and (\ref{eq:c0largeN}), one can 
verify that the contributions to
the energy per particle from Eq.~(\ref{eq:eden-1}) are indeed 
of ${\cal O}(1/g)$. 
To this order, similar RPA contributions appear in the modified
loop expansion discussed in Refs.~\cite{WEISS83} and
\cite{WEHRBERGER90}.
We will calculate $\widetilde{\cal E}_{NLO}$
numerically and discuss the energy density below.
In the next section we comment on the higher-order
contributions.

\subsection{Higher-order Effective Potential}

The higher-order contributions come from connected
1PI-diagrams built from the vertices in Eq.~(\ref{eq:etavertc0})
and the $\sigma$ propagator
\beq
D_\sigma (q)=\frac{i}{C_0}\left(1+
ig\, C_0\,\int\frac{d^4 p}{(2\pi)^4} G_H(p+q) G_H(p)\right)^{-1}
=\frac{i}{C_0}\frac{1}{1-C_0 \Pi_0(q)}\,.
\label{eq:dsig}
\eeq
The diagrams occuring at two-loops are shown in Fig.~\ref{fig_dian2lo}.
Using Eqs.~(\ref{eq:c0largeN}) and (\ref{eq:dsig})
it is easy to see that $D_\sigma \sim {\cal O}(g)$.
It is then straightforward to verify 
that the 1PI-diagrams shown in Fig.~\ref{fig_dian2lo}
are of ${\cal O}(1/g^2 )$ in the energy per particle. 
In fact, these are the only contributions
at this order, with all other diagrams being suppressed by at least one
more power of $1/g$. A general argument can be made as follows:\footnote{
The argument does not apply in the present form if many-body forces are 
present. However, we will use a renormalization argument below to 
identify at which order three-body forces contribute.}
Fermion loops and $\sigma$ propagators contribute a positive power 
of $g$ while a $\sigma$-fermion-fermion vertex is proportional to $C_0$
and contributes a negative power. The contribution
to the energy per particle of diagram with $L_F$ fermion loops, $V$ 
$\sigma$-fermion-fermion vertices, and $I_\sigma$ $\sigma$ propagators
scales as $g^\nu$ with
\beq
\nu=L_F-V+I_\sigma-1\,.
\label{eq:nu1}
\eeq
We define the number of $\sigma$ loops as the total number of loops
minus the number of fermion loops $L_\sigma \equiv L-L_F$. 
The number of $\sigma$ propagators must be half the number
of $\sigma$-fermion-fermion vertices $I_\sigma=V/2$, and  twice
the total number of internal lines is equal to three times the number
of $\sigma$-fermion-fermion vertices, $2I=3V$.
Using the general topological relation relating the total number of
loops $L$, vertices $V$, and internal lines $I$ of a given diagram:
$L=I-V+1$, we can rewrite Eq.~(\ref{eq:nu1}) as
\beq
\nu=L_F - I_\sigma -1 = -(L-L_F) =  -L_\sigma\,,
\label{eq:nu2}
\eeq
which proves that the loop-expansion of the effective action is
equivalent to a $1/g$ expansion of the energy per particle
if no many-body forces are present.

An important question is at what order in $1/g$ three-body
forces enter. In Appendix \ref{sec:appho}, we discuss how many-body forces can
be included in our formalism. Here we use a renormalization argument 
to show that they will contribute at ${\cal O}(1/g^2 )$.
\begin{figure}[t]
\centerline{\includegraphics*[width=12cm,angle=0]{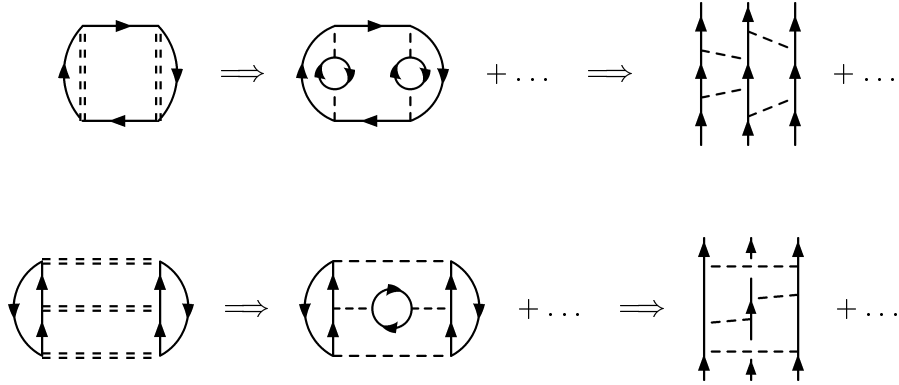}}
\vspace*{-.0in}
\caption{Diagrams with logarithmically divergent contributions to the
effective action at two loops.
The solid lines indicate Hartree propagators $G_H$, the double dashed 
lines indicate the full $\sigma$ propagator $D_\sigma$, and 
the single dashed lines indicate the bare $\sigma$ propagator
(a single potential exchange).
}
\label{fig_log_div}
\end{figure}        
In Fig.~\ref{fig_log_div}, we illustrate two particular contributions
that are contained in the two-loop diagrams from
Fig.~\ref{fig_dian2lo} by expanding the $\sigma$ propagator.
These two diagrams have logarithmic UV divergences. 
This can be seen most easily by cutting the hole lines to obtain
the corresponding diagram in free space. The resulting diagrams
are 1-particle irreducible and describe three-particle scattering.
They have the same UV divergences as the finite-density diagrams in 
Fig.~\ref{fig_log_div}.  Proper
renormalization of this UV divergence requires a three-body 
contact interaction of the type $(\psi^\dagger \psi)^3$. 
This was first noticed for a system of bosons in Ref.~\cite{Braaten97} and 
later applied to fermions in Ref.~\cite{HAMMER00}.
For the $1/g$ expansion to be consistent,
the three-body counterterm must appear at the same order as
the diagrams it renormalizes, so
three-body forces have to enter at
${\cal O}(1/g^2)$.\
This is manifest in the renormalization-group equation for the running
of the non-derivative contact three-body force $D_0$ (see
Ref.~\cite{HAMMER00}), which depends on $C_0^4$.

\section{Energy Density and Stability Analysis}
\label{sec:results}

\subsection{Calculation of $\widetilde{\cal E}_{NLO}$}
We now turn to the calculation of $\widetilde{\cal E}_{NLO}$.
In order to evaluate the integral in Eq.~(\ref{eq:ed1}), we
need the polarization insertion $\Pi_0$.
The calculation of the polarization insertion is standard and
can be found, for example, in  
Ref.~\cite{FETTER71}. Here we only quote the final result:
\bea
{\rm Re}\, \Pi_0(v_0,v)&=&\frac{gM\kf}{4\pi^2}
\bigg\{1+\frac{1}{2v}
 \left(
   1 - \left(v_0/v - v/2\right)^2
 \right)
 \ln\left|\frac{1+ (v_0/v-v/2)}{1-(v_0/v-v/2)}\right| \nonumber\\
&& -\frac{1}{2v}
 \left(
   1 - \left(v_0/v + v/2\right)^2
 \right)
 \ln\left|\frac{1+ (v_0/v+v/2)}{1-(v_0/v+v/2)}\right|\bigg\} 
\label{eq:repi} \\[5pt]
{\rm Im}\, \Pi_0(v_0,v)&=&-\frac{gM\kf}{8\pi v} \left(
 1-\left(v_0/v - v/2\right)^2 \right)\,;\qquad
 v>2,\; \frac{v^2}{2}+v \geq v_0 \geq \frac{v^2}{2}-v\,, \nonumber\\
&=&-\frac{gM\kf}{8\pi v} \left(
 1-\left(v_0/v - v/2\right)^2 \right)\,;\qquad
 v<2,\; v+\frac{v^2}{2} \geq v_0 \geq v-\frac{v^2}{2}\,, \nonumber\\
&=&-\frac{gM\kf}{8\pi v} 2v_0\,;\qquad
 v<2,\; 0 \leq v_0 \leq v-\frac{v^2}{2}\,,\nonumber\\
&=& 0\,; \qquad\mbox{otherwise} \,.
\label{eq:impi}
\eea

\begin{figure}[t]
\centerline{\includegraphics*[width=10cm,angle=0]{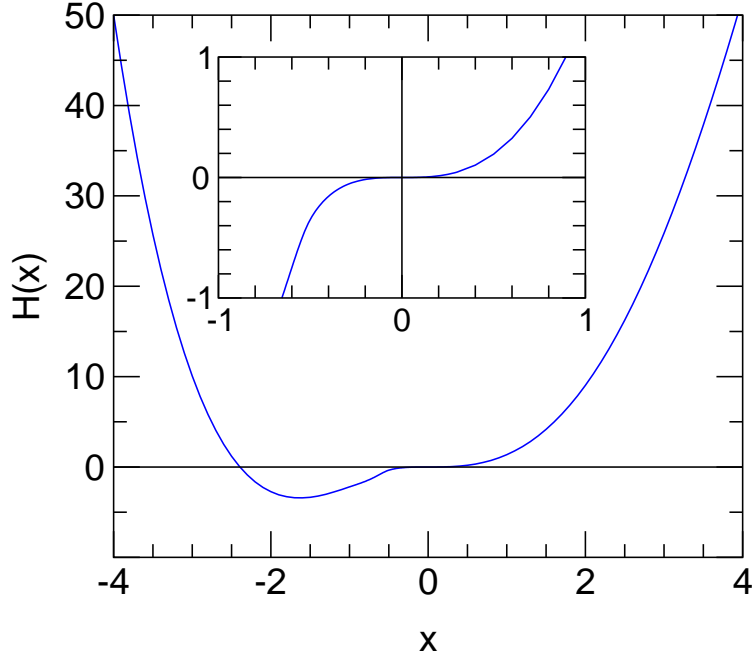}}
%\vspace*{-.2in}
\caption{The universal function $H(x)$ for $-4 \leq x \leq 4$. 
The solid line shows the result of the numerical
calculation. The inset shows $H(x)$  for $-1 \leq x \leq 1$
in more detail.}
\label{Hxplo}
\end{figure}

For the calculation of $\widetilde{\cal E}_{NLO}$ it is convenient
to define the dimensionless variable
\beq
x\equiv\frac{gM\kf}{4\pi^2}C_0=\frac{g}{\pi}\kf a_s\,,
\eeq
and the universal function
\beq
F_0(v_0,v)\equiv \frac{gM\kf}{4\pi^2} \Pi_0(v_0,v)\,.
\eeq
The integral in Eq.~(\ref{eq:ed1}) can then be written as
\bea
\widetilde{\cal E}_{NLO} &=& \frac{\kf^5}{4\pi^3 M}
\int_0^\infty  v^2 dv  \int_0^\infty dv_0 \Bigg[
x\, {\rm Im}\, F_0(v_0,v)+x^2\, {\rm Im}\, F_0(v_0,v)
{\rm Re}\, F_0(v_0,v) \nonumber\\
&& \qquad\qquad\qquad\qquad\qquad\qquad -\arctan \left(\frac{
x\, {\rm Im}\,F_0(v_0,v)}{1-x\, {\rm Re}\,F_0(v_0,v)}\right)
\Bigg]\nonumber\\
&\equiv& \frac{\kf^5}{4\pi^3 M} H(x)
\label{eq:ed1b}
\eea
where $H(x)$ is a universal function of $x$ alone.
Due to the complicated form of $\Pi_0(v_0,v)$, Eq.~(\ref{eq:ed1b})
has not been evaluated analytically. Since $\widetilde{\cal E}_{NLO}$
is manifestly finite, however, $H(x)$ can be calculated numerically.
In Fig.~\ref{Hxplo}, we show the results
of a numerical evaluation of $H(x)$. 
The inset shows $H(x)$ in the perturbative region $-1 \leq x \leq 1$
in more detail. The local minimum for negative $x$ allows for
a self-bound uniform system. 
(Repulsion at NLO was also observed in the modified loop expansion
of Ref.~\cite{WEHRBERGER90}.)

\begin{figure}[t]
\centerline{\includegraphics*[width=12cm,angle=0]{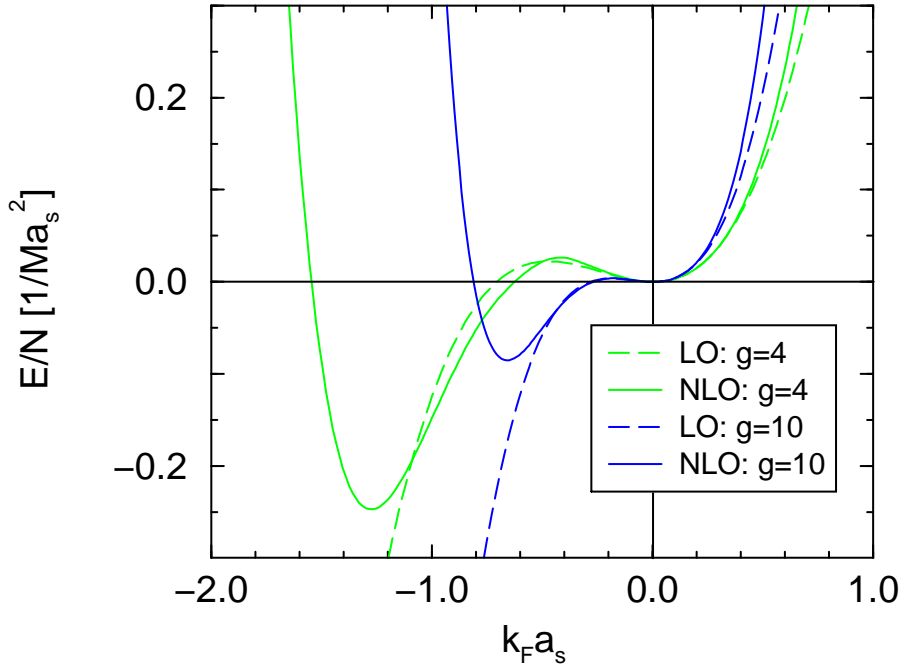}}
\vspace*{-.2in}
\caption{Energy per particle in units of $1/Ma_s^2$ at leading order (LO) and
next-to-leading-order (NLO) in the $1/g$ expansion, for $g=4$ and $g=10$.}
\label{EoNglonlo}
\end{figure}

The energy per particle through NLO can be written in terms of $g$ and $x$
as
\beq
  \frac{E}{N} = \frac{\kf^2}{2M}
  \left[
    \left( \frac{3}{5} + \frac{2}{3}x \right)
    + \frac{1}{g} \left(
    \frac{3}{\pi}H(x) - \frac{2}{3}x + \frac{4}{35}(11 - 2\ln 2)x^2
    \right)
  \right] \ .
\label{eq:eonx}
\eeq
In Fig.~\ref{EoNglonlo}, we show the energy per particle in units of the 
Fermi energy for two different values of $g$: $g=4$ and $g=10$.
The dashed lines show the LO results while the solid
lines show the NLO order results. The difference between
LO and NLO gives an indication of where the expansion
breaks down. The NLO result oscillates around the LO result in the
vicinity of $\kf a_s=0$ and grows without limit as $\kf a_s \to\pm \infty$.
As $g$ increases, the minimum becomes
shallower and moves toward smaller values of $\kf a_s$.

An important consideration is whether the energy calculated above actually 
corresponds to the true ground state of the system.
There are two issues here: whether the state that evolves
adiabatically from the noninteracting ground state is unstable to
pairing near the Fermi surface and whether a homogeneous system is stable.
There is no pairing instability if we restrict ourselves to repulsive
interactions (e.g., a positive, natural-size scattering length), but
pairing is inevitable with an attractive interaction (e.g., a negative
scattering length or $x<0$).  
We will investigate the role of pairing in the EFT framework in future
work.  We note here that more conventional investigations typically
find that the impact of pairing on the bulk properties (on which we
focus) is small for natural scattering lengths \cite{Papenbrock:1998wb}.

In order to investigate the stability of the homogeneous state, we
calculate the compressibility \cite{JACKSON94}
\bea
K&=&\frac{\partial}{\partial n} \left(n^2 \frac{\partial}{\partial n} 
    \frac{E}{N} \right)\,.
    \label{eq:compress}
\eea
When the compressibility is positive, the system is stable with respect
to infinitesimal long-range fluctuations in the density. When the
compressibility is negative, however, long-range fluctuations will grow
exponentially. As a consequence, the system will cluster and the
homogeneous state is not the ground state.
Using Eq.~(\ref{eq:eden0}), the compressibility at leading 
order in the $1/g$ expansion is
\beq
K_{LO}=\frac{2\kf^2}{3M}\left(\frac{g\kf a_s}{\pi}+\frac{1}{2}\right)\,,
\eeq
which turns negative for
\beq
  x < x_{\rm crit} =-\frac{1}{2}\,.
\label{eq:xcrit}
\eeq
As a consequence, the homogeneous system is unstable if
$x < -1/2$. This constraint was previously obtained in Ref.~\cite{JACKSON94}
by identifying the radius of convergence for the exact fermion RPA energy, 
as well as from a variational argument. At next-to-leading order 
in the $1/g$ expansion, the compressibility $K_{NLO}$
follows from Eqs.~(\ref{eq:eonx}) and (\ref{eq:compress}) 
but we have not obtained an
analytic expression.
Furthermore, the point where the compressibility becomes negative 
now depends on $g$ and $\kf a_s$ separately. Nevertheless, the numerical 
results for $K_{NLO}$ are close to the leading order values (\ref{eq:xcrit}). 
For example, the next-to-leading order compressibility turns negative 
if $\kf a_s < -0.73, -0.37, -0.15$ for $g=2,4,10$, respectively, as 
compared to $\kf a_s < -0.79, -0.39, -0.16$ at leading order.
This means that the self-bound uniform state given by the minimum
of $E/N$ in Fig.~\ref{EoNglonlo} is not a stable ground state.

The above constraints can also be obtained directly from the 
effective action by looking for minima in Eq.~(\ref{eq:effactapp})
corresponding to a nonuniform state (with $\sigma_c=\sigma_c(x)$).
The saddle points of the effective action determine where such
minima appear. This analysis also determines the density fluctuation
modes that destabilize the homogeneous state.
In leading order, we require:
\beq
0=\frac{\delta^2}{\delta\sigma_c(x)\delta\sigma_c(y)}\Gamma_{LO}[\sigma_c(z)]=
\frac{\delta^2}{\delta\sigma_c(x)\delta\sigma_c(y)}\left[
\frac{g}{i}\,\Tr\ln[G_H^{-1}(x,y)]+\frac{C_0}{2}
\int\!d^4x\,\sigma_c(x)^2\right]\,,
\eeq
where $\sigma_c$ is a function of $x$. Using the identity
\beq
\frac{\delta^2}{\delta\sigma_c(x)\delta\sigma_c(y)}\Tr\ln[G_H^{-1}]=
-C_0^2 G_H(x,y) G_H(y,x)\,,
\eeq
performing a Fourier transform with respect to $x-y$, and setting
$q_0=0$, we obtain the equation
\beq
0=1-C_0\Pi_0 (q_0=0,q)=
1+x\left\{1-\frac{\kf}{q}\left(1-\frac{q^2}{4\kf^2}\right)
\ln\left|\frac{1-q/(2\kf)}{1+q/(2\kf)}\right|\right\}\,,
\label{eq:stabana}
\eeq
whose solutions determine the long-range density fluctuation modes that
destabilize the homogeneous system. The constraint obtained from solving
Eq.~(\ref{eq:stabana}) agrees with Eq.~(\ref{eq:xcrit}).
At this order, the stability analysis in the effective action formalism 
is equivalent to identifying the poles in the $\sigma$ propagator
$D_\sigma(q_0=0,q)$ from Eq.~(\ref{eq:dsig}), as is done in conventional
many-body approaches \cite{FETTER71,NEGELE88}. In principle,
it is straightforward to extend
the stability analysis to next-to-leading order by requiring
\beq
0=\frac{\delta^2}{\delta\sigma_c(x)\delta\sigma_c(y)}
\left\{\Gamma_{LO}[\sigma_c(z)]+\Gamma_{NLO}[\sigma_c(z)]\right\}\,.
\eeq
However, this analysis involves the evaluation of two-loop diagrams 
and is beyond the scope of this work.

\section{Bose limit}
\label{sec:bose}

The EFT expansion is not limited, in general, to small values of $x$.
For example, Eq.~(\ref{eq:eonx}) implies that the $1/g$ expansion
is valid for large $x$ if $g$ is also sufficiently large.
The Bose limit, in which $x \rightarrow \infty$, provides an
instructive example.

The energy density of a Bose system can be obtained from the 
energy density of a Fermi system by taking the limit $g\to \infty$, 
$\kf \to 0$, but keeping $n=g\kf^3/(6\pi^2)$ constant~\cite{JACKSON94}.
Here $g$ is the degeneracy of an artificial ``flavor'' quantum number.
This can be understood intuitively starting from a noninteracting
Fermi system with large degeneracy.
If the degeneracy $g$ is greater than the number of particles,
the spatial state will be a symmetric wave function with all particles
in the lowest momentum state.  This is the same as the wave function
of the Bose system with the same number of particles.
If the interacting state evolves adiabatically from the noninteracting
state, then the limiting Fermi system is the same as the interacting
Bose system times a totally antisymmetric
``flavor'' wave function that has no physical
consequences \cite{BRANDOW71}.
 
The Bose limit can be taken directly in Eq.~(\ref{eq:edkf7}),
leading to the mean-field energy for a dilute Bose system. 
All terms vanish
except for the part of the second term proportional to $g$,
and the result is 
\beq
{\cal E}_0^{B}=\frac{2\pi a_s n^2}{M}\,,
\eeq
in agreement with the literature \cite{FETTER71}.
For higher-order terms, the analysis is more complicated since
some contributions diverge as $g \to \infty$ and therefore need to be resummed.
However, this is precisely the same resummation that is implemented in the 
one-loop effective action from the previous section. 

In the Bose limit,
the polarization insertion $\Pi_0$ [see Eq.~(\ref{eq:polins})] 
has a simple analytic expression.
In order to evaluate Eq.~(\ref{eq:effactapp1}) most easily, we take
the Bose limit of $\Pi_0$ before performing the $dy$ and $d^4 q$ integrals.
The polarization insertion can be written as
\beq
\Pi_0(q_0,\vec{q})=g\int\frac{d^3 p}{(2\pi)^3}\theta(\kf-p) \left(
\frac{\theta(|\vec{p}+\vec{q}|-\kf)}{q_0+\omega_\vec{p}-\omega_{\vec{p}
+\vec{q}}+i\epsilon}-\frac{\theta(|\vec{p}-\vec{q}|-\kf)}{q_0-\omega_\vec{p}
+\omega_{\vec{p}-\vec{q}}-i\epsilon}\right)\,.
\eeq
If we let $|\vec{q}|=q \gg \kf$ (or equivalently take $\kf\to  
0$), the $\theta$ functions involving $\vec{q}$ can be dropped. 
Expanding
\beq
\omega_\vec{p}-\omega_{\vec{p}\pm \vec{q}}=
-\omega_\vec{q} (1+{\cal O}(p/q))\,,
\eeq
the polarization insertion in the Bose limit becomes \cite{JACKSON94}
\bea
\Pi_0^B(q_0,\vec{q})&=&g\int\frac{d^3 p}{(2\pi)^3} \theta(\kf-p) \left(
\frac{1}{q_0-\omega_\vec{q}+i\epsilon}-\frac{1}{q_0+\omega_\vec{q}
-i\epsilon}\right)\nonumber\\
&=&\frac{2\omega_\vec{q} n}{(q_0-\omega_\vec{q}+i\epsilon)
(q_0+\omega_\vec{q}-i\epsilon)}\,.
\eea
Thus $\Pi_0^B$ has simple poles in $q_0$ given by the noninteracting 
single-particle kinetic energy.

The ring sum leading to the integrand in Eq.~(\ref{eq:effactapp1})
shifts the poles to the Bogoliubov quasiparticle energies:
\beq
  \frac{y \Pi_0(q)}{1 - y\Pi_0(q)} =
  \frac{y}{\Pi_0^{-1}(q) - y}
  = \frac{2\omega_\vec{q} n y}{q_0^2 - \omega_\vec{q}^2
    - 2 \omega_\vec{q} n y + i\epsilon}
    = \frac{2\omega_\vec{q} n y}{q_0^2 - \epsilon_\vec{q}^2
      + i\epsilon} \ ,
\eeq
where 
\beq
   \epsilon_\vec{q} \equiv \sqrt{\omega_\vec{q}^2 
     + 2 \omega_\vec{q} n y} \ .
\eeq
The $dq_0$ integral in
Eq.~(\ref{eq:effactapp1}) is simply evaluated
by contour integration and leads to
\beq
{\cal E}_1^B = -\frac{\Gamma_{NLO}}{VT} =
\frac{n}{2} \int_0^{C_0}\! dy \int\!\frac{d^3 q}{(2\pi)^3}
\frac{\omega_\vec{q}}{\epsilon_\vec{q}}
=
\frac{n}{2} \int_0^{C_0}\! dy \int\!\frac{d^3 q}{(2\pi)^3}
\frac{q^2}{q\sqrt{q^2+4Mny}}\ .
  \label{eq:E1B}
\eeq
Note that an expansion  of the final integrand in powers of $y$ would 
generate infrared (IR) divergences. The resummation of the one-loop fermion
effective action is therefore required for bosons even in the 
low-density limit.
This summation of ring diagrams because of IR divergences is
analogous to the calculation of the correlation energy for a uniform 
electron gas, only in that case it is the interaction that is the source
of the IR divergences \cite{FETTER71}.
 
The $d^3 q$ integral in Eq.~(\ref{eq:E1B})
can be evaluated in dimensional regularization
with minimal subtraction.
In particular, the formula
\beq
   \int\! \frac{d^dk}{(2\pi)^d} \,
   \frac{k^2}{k\sqrt{k^2 + \Lambda^2}}
   =
  \frac{\Gamma(-d/2)\Gamma(1/2+d/2)}%
     {(4\pi)^{d/2}\Gamma(d/2)\Gamma(1/2)} 
   (\Lambda^2)^{d/2}
\eeq
can be derived and applied for $d=3$ \cite{PESKIN95}.
The end result is
\bea
{\cal E}_1^B &=& \frac{n}{15\pi^2} (4Mn)^{3/2} C_0^{5/2}=
\frac{2\pi a_s n^2}{M}\; \frac{128}{15\sqrt{\pi}}\sqrt{na_s^3}\,,
\eea
in agreement with the well-known result \cite{FETTER71}.\footnote{If 
one starts from the subtracted expression (\ref{eq:effactapp1b}) instead
of Eq.~(\ref{eq:effactapp1}), the same result is obtained
with a direct integration. The
separately calculated terms proportional to $\kf a_s$ and $(\kf a_s)^2$
in Eq.~(\ref{eq:eden-1})
vanish in the Bose limit, while the subtractions implicit in 
Eq.~(\ref{eq:effactapp1b}) cancel linear and cubic UV divergences.
If dimensional regularization with minimal subtraction is used as above,
these UV divergences are subtracted automatically and one can use 
Eq.~(\ref{eq:effactapp1}) directly.}

In principle, our numerical evaluation of the universal
function $H(x)$, illustrated in 
Fig.~\ref{Hxplo}, should match on to the Bose limit for large,
positive $x \simgt 10$.  
We have not been able to demonstrate this matching.  
Indeed, the numerical continuation to large $x$ of $H(x)$ exhibits
an entirely different behavior (e.g., it reaches a maximum and then
turns negative, going to minus infinity with a different asymptotic
power of $x$ then predicted by the Bose limit).  
It is not clear whether the problem is one of numerics (our calculations
$x \simgt 10$ show severe round-off errors) or a more fundamental problem.
Further investigations to resolve this issue are in progress.

\section{Summary}
\label{sec:summary}

In this paper,
a system of fermions with short-range interactions at finite density 
is studied using the framework of effective field theory.
The effective action formalism for fermions with auxiliary fields 
leads to a loop expansion in which particle-hole bubbles are resummed to
all orders.  For spin-independent interactions, the loop expansion is 
equivalent to a systematic expansion in $1/N$,
where ``$N$'' is the spin-isospin degeneracy $g$.
Thus we have a double expansion, in $1/g$ as well as $\kf a_s$. 
The loop expansion differs from the dilute expansion described in
Ref.~\cite{HAMMER00}, even at leading order.
At next-to-leading order, the expansion requires Hartree plus RPA
contributions to the ground state \cite{FUKUDA94,WEHRBERGER90,TANAKA93}.

The formalism enables us to examine uniform systems for potential
self-bound solutions.
At next-to-leading order in $1/g$, the nonperturbative resummation
leads for the uniform system
to a non-trivial and non-analytic dependence on the Fermi momentum
$\kf$.  
There is a self-bound minimum within the radius of convergence
of the EFT for sufficiently large $g$, but a stability analysis
reveals that it is unstable to
density fluctuations (e.g., clustering).  

An interesting limit of the large $g$ expansion takes $g$ to infinity
and $\kf$ to zero, with the density $n$ held fixed.
This limit reproduces the energy expansion for a dilute Bose gas,
with the non-analytic dependence on $\sqrt{n}$ generated by the
infinite summation of ring diagrams.
We have not succeeded in reproducing the Bose limit starting from
our explicit calculations for fermions,
which may simply reflect the numerical difficulties in carrying out
the limit but might also mean that there are subtleties we have not
recognized.  Work is in progress to resolve this issue.

While the present investigation illustrates some features of a
systematic nonperturbative treatment of a Fermi system, we need to
extend our treatment to include pairing and large scattering lengths
to make contact with the physical systems of greatest interest.
Investigations in these areas are in progress as well as work to adapt
the effective action formalism to finite systems in the form of
density functional theory \cite{PUGLIA02}.

\acknowledgments

We thank  P.~Bedaque, S.~Jeschonnek, S.~Puglia, A.~Schwenk, and B.~Serot 
for useful comments.
HWH thanks the Benasque Center for Science for its hospitality and
partial support during completion of this work. 
This work was supported in part by the National Science Foundation
under Grant No.~PHY--0098645.

\begin{appendix}

\section{Higher-Order Two-Body and Many-Body Forces}
\label{sec:appho}

In this appendix, we generalize the discussion of the $1/N$ expansion
to consider higher-order
two-body interactions ($C_2$, $C'_2$, and so on) as well as many-body
forces ($D_0$ and so on) that are present in a general EFT Lagrangian for the
dilute system (see Ref.~\cite{HAMMER00} for definitions and details):
\bea
  {\cal L}  &=&
       \psi^\dagger \biggl[i\partial_t + \frac{\nab^{\,2}}{2M}\biggr]
                 \psi - \frac{C_0}{2}(\psi^\dagger \psi)^2
            + \frac{C_2}{16}\Bigl[ (\psi\psi)^\dagger 
                                  (\psi\galnab^2\psi)+\mbox{ h.c.} 
                             \Bigr]   
  \nonumber \\[5pt]
   & & \null +
         \frac{C_2'}{8} (\psi \galnab \psi)^\dagger \cdot
             (\psi\galnab \psi) 
         - \frac{D_0}{6}(\psi^\dagger \psi)^3 +  \ldots\,.
  \label{applag}
\eea
There are various different ways of incorporating these interactions
into the loop expansion, which correspond to different reorganizations
of the original perturbative expansion, with different resummations of
diagrams. The details of the system under consideration
determine the power counting of the
higher-order contributions; the most appropriate resummation will
implement that power counting.

At the simplest level, we can incorporate all higher-order vertices
perturbatively.
This means that the only infinite summations based on large $g$
will be of the same subdiagrams with $C_0$ vertices we have already
considered. 
We return to the generating functional in Eq.~(\ref{eq:Zorig}),
but now with the full EFT Lagrangian for a dilute system  with
short-range interactions, and with Grassmann sources $\xi^\dagger$
and $\xi$ coupled to the fermion fields:  
\beq
Z[\xi^\dagger, \xi]=e^{-iE[\xi^\dagger, \xi]}=\int{\cal D}\psid{\cal D}\psi\,
e^{i\int\! d^4 x \,[{\cal L} + \xi^\dagger(x)\psi(x) 
    + \psi^\dagger(x)\xi(x)]}\,,
  \label{eq:Zxi}
\eeq
where ${\cal L}$ is given by Eq.~(\ref{eq:lag}).
[Spin indices are implicit in Eq.~(\ref{eq:Zxi}).]
Functional derivatives with respect to the Grassmann sources can be used
to generate the $n$-point fermion Green's functions.  
If we isolate the quadratic part of the Lagrangian, this procedure
generates the perturbative expansion of these functions
in terms of the noninteracting
fermion propagator (at finite chemical potential).
The sum of closed,
connected diagrams generated this way reproduces the perturbative
expansion of the energy density from Ref.~\cite{HAMMER00}.

In the present context, we use the Grassmann sources to remove all
interaction terms from the path integral ${\cal L}_I(\psi^\dagger,\psi)
\rightarrow {\cal L}_I(-i\delta/\delta \xi, -i\delta/\delta\xi^\dagger)$
\emph{except} for the $C_0$ contact interaction. 
To carry out the procedure in Sect.~\ref{sec:auxiliary}
leading from Eq.~(\ref{eq:Zorig}) to Eq.~(\ref{eq:ef1}),  
we need to incorporate the new
Grassmann source terms.  This is straightforward, since the identity
\beq
  \psi^\dagger G^{-1}\psi + \psi^\dagger\xi + \xi^\dagger\psi
  =
  (\psi^\dagger + \xi^\dagger G)G^{-1}(\psi + G\xi)
     - \xi^\dagger G \xi\,,
\eeq
and a shift in $\psi$ and $\psi^\dagger$ leaves the same integral as
before (leading to the same $\Tr\ln$ term) \emph{plus} 
a new term:
\beq
  \exp\left[- i \int\!d^4x\,d^4y\, \xi^\dagger(x) G(x,y) \xi(y)\right]
  \, .
  \label{eq:newterm}
\eeq
The new term still depends on $\sigma$ through $G(x,y)$.

\begin{figure}[t]
\centerline{\includegraphics*[width=8cm,angle=0]{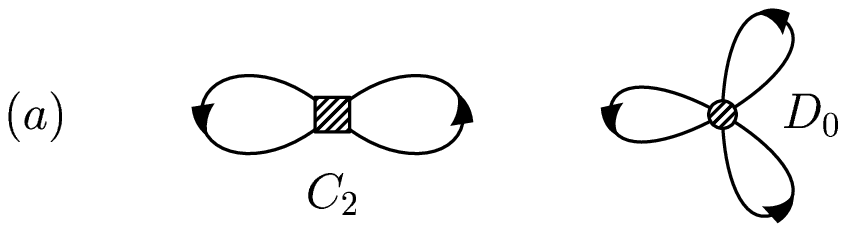}}
\vspace*{+.2in}
\centerline{\includegraphics*[width=12cm,angle=0]{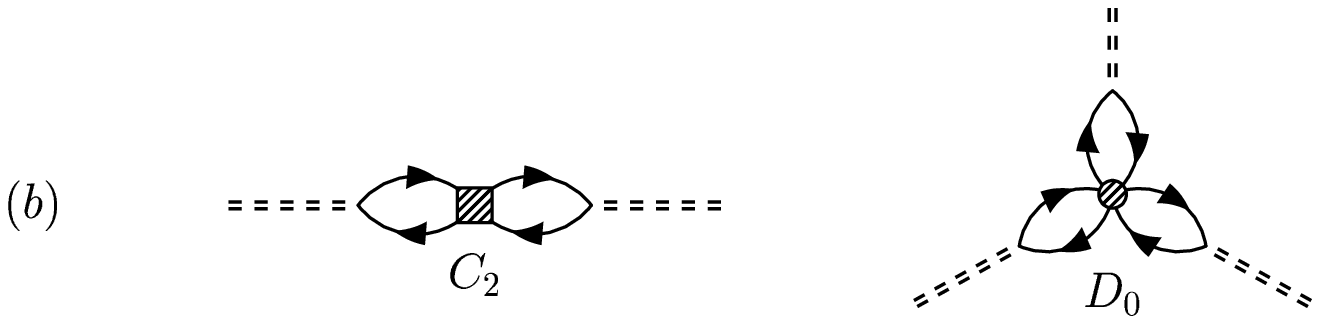}}
\caption{(a) New diagrams and (b) $\sigma$ propagator and vertex corrections
generated by the perturbative inclusion of higher-order terms
in the effective Lagrangian (\ref{applag}).}
\label{fig:C2D0}
\end{figure}        

As in Eq.~(\ref{eq:effact}), we expand $\sigma$ about $\sigma_c$ in 
Eq.~(\ref{eq:newterm}), which has the effect of expanding $G$ about $G_H$.
The Grassmann sources $\xi$ and $\xi^\dagger$ in Eq.~(\ref{eq:newterm})
have to match up with the derivatives in 
${\cal L}_I(-i\delta/\delta \xi, -i\delta/\delta\xi^\dagger)$
because the sources
 are set to zero in the end.
This dictates how the higher-order vertices (e.g., $C_2$ and $D_0$)
appear in Feynman diagrams attached to $G_H$ propagators.
The quadratic term in $\eta$ can be directly incorporated into
Eq.~(\ref{eq:spropc0}) while the higher-order terms generate additional
connected 1PI-diagrams.  (For consistency, one should expand the
$\sigma$ propagators perturbatively in the higher-order interactions.)
The $\eta$-independent part of Eq.~(\ref{eq:newterm}), which has
$G \rightarrow G_H$, generates tadpole-like diagrams with the new
vertices, such as those illustrated in Fig.~\ref{fig:C2D0}(a),
as well as higher-order contributions. 
The new diagrams based on the $\eta$ expansion can be constructed by breaking
every possible $G_H$ line (including those buried in the $\sigma$
propagator) and connecting the broken ends to the new vertices.
Examples of $\sigma$-propagator and $V_3$ vertex corrections
are shown in Fig.~\ref{fig:C2D0}(b).     

An alternative approach to incorporating higher-order two-body and
many-body interactions is to generalize the $\sigma$ Lagrangian
in Eq.~(\ref{eq:genfun}) to include $\sigma^3$ and higher-order
self-interactions, as well as terms with gradients (but not time
derivatives) acting on $\sigma$. 
If we put back the fermion path integral and eliminate the $\sigma$
field by iteratively applying its Euler-Lagrange equation via field
redefinitions, we reproduce an infinite subset of the terms in the
general dilute effective Lagrangian. 
For example, we recover 
the $D_0$ vertex and the combination of $C_2$ and $C'_2$
corresponding to a term proportional to
$\nabla(\psi^\dagger\psi)\cdot\nabla(\psi^\dagger\psi)$.
In general, we find all interaction terms with
$\psi^\dagger_\alpha\psi_\alpha$ to some power and gradients acting on
such factors
(these are terms
that depend only on the momentum \emph{transfer} between the fermions).

The generalized $\sigma$ Lagrangian can be expanded about $\sigma_c$ as
before, with the new terms treated to all orders (analogous to the
treatment of the linear sigma model in Ref.~\cite{PESKIN95}) or
as perturbative corrections.  In the former case, the strict power
counting in which all diagrams contribute at a given order in
$1/g$ is violated unless the counting of the coefficient of the $\sigma^3$ 
term, $\widetilde{D}_0 \propto D_0$, is promoted to ${\cal O}(1/g^2)$ from
${\cal O}(1/g^4)$ in the perturbative case.  
The latter case is similar to the discussion
above, except that a given higher-order term in the generalized $\sigma$
Lagrangian will be equivalent to a selective summation of higher-order
terms expressed in the basis of fermion fields alone.

\end{appendix}

\end{document}